\documentclass[%
 aip,
 sd,%
 amsmath,amssymb,
twocolumn,
 tightenlines,
 10pt 
]{revtex4-1}

\usepackage{graphicx}
\usepackage{dcolumn}
\usepackage{bm}
\usepackage{hyperref}

\newcommand{\unit}[1]{\ensuremath{\, \mathrm{#1}}}

\begin{document}


\title[Akkus \textit{et al.}]{ \large Atomic Scale Interfacial Transport at an Extended Evaporating Meniscus}

\normalsize

\author{Yigit Akkus}
 
\affiliation{\mbox{Lyle School of Engineering, Southern Methodist University, Dallas, TX 75205, USA}}%
 
\affiliation{ASELSAN Inc., 06172 Yenimahalle, Ankara, Turkey
}%

\author{Anil Koklu}  
\author{Ali Beskok}
\email{abeskok@smu.edu}
\affiliation{\mbox{Lyle School of Engineering, Southern Methodist University, Dallas, TX 75205, USA}}%

\date{\today}

\begin{abstract}
Recent developments in fabrication techniques enabled the production of nano- and \r{a}ngstr\"{o}m-scale conduits. While scientists are able to conduct experimental studies to demonstrate extreme evaporation rates from these capillaries, theoretical modeling of evaporation from a few nanometers or sub-nanometer meniscus interfaces, where adsorbed film, transition film and intrinsic region are intertwined, is absent in the literature. Using the computational setup constructed to identify the detailed profile of a nano-scale evaporating interface, we discovered the existence of lateral momentum transport within and associated net evaporation from adsorbed liquid layers, which are long believed to be at the equilibrium established between equal rates of evaporation and condensation. Contribution of evaporation from the adsorbed layer increases the effective evaporation area, reducing the excessively estimated evaporation flux values. This work takes the first step towards a comprehensive understanding of atomic/molecular scale interfacial transport at extended evaporating menisci. The modeling strategy used in this study opens an opportunity for computational experimentation of steady-state evaporation and condensation at liquid/vapor interfaces located in capillary nano-conduits.
%
\end{abstract}


\maketitle

\begin{figure} [h!] 
\setlength{\belowcaptionskip}{-15pt} 
\includegraphics[width=3.4in]{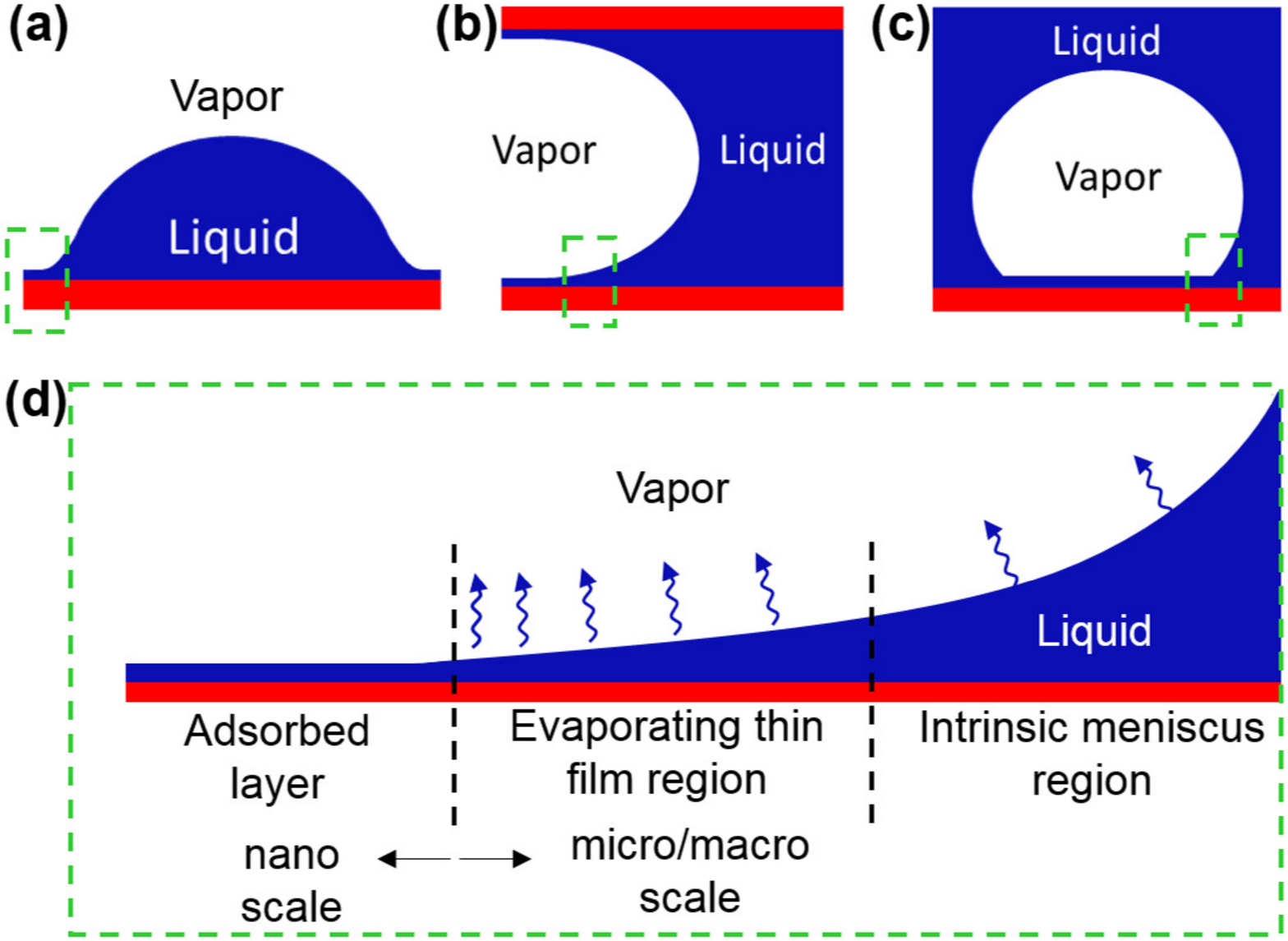}
\caption{\label{fig:fig1} Schematics of (a)~an evaporating droplet, (b)~evaporation from a capillary and (c)~bubble growth. Green dashed rectangles point out the region near contact line, where evaporation rate intensifies due to the decreased liquid film thickness. (d)~Liquid forms three multiscale regions around the contact line. Film thickness is at the macro-scale within intrinsic meniscus region. Capillary pressure gradient, arising from the curvature changes, drive the flow in this bulk region. Film thickness is at the micro-scale within evaporating thin film region, where evaporation rate intensifies due to the reduced film resistance. Changes in capillary and disjoining pressures govern the liquid flow in this region. Film thickness is at the nano-scale within adsorbed layer. }
\end{figure}

Capillary evaporation and the associated passive liquid flow are vital for numerous natural and artificial processes such as transpiration of water in plants,\cite{wheeler2008} solar steam generation, \cite{ghasemi2014,ni2016} water desalination,\cite{lee2014} microfluidic pumping,\cite{lynn2009} and cooling of electronic and photonic devices. \cite{li2012enhancing} Regardless of the process or the geometrical configuration, studies on evaporation focus on identification and characterization of heat transfer and flow dynamics in the vicinity of the contact line, the juncture of three phases of matter. \mbox{Fig.~\ref{fig:fig1}a-c} shows different evaporation processes schematically. Green dashed rectangles point out the liquid film distribution around the contact line, which is broadly composed of three multiscale regions as shown in \mbox{Fig.~\ref{fig:fig1}d}. Evaporation rate intensifies in evaporating thin film region due to the micro-scale liquid film thickness. The adsorbed nano-scale layer extending further is assumed to be non-evaporating due to the suppression of evaporation by strong long-range intermolecular forces.\cite{wayner1976,truong1987,wayner1991,dasgupta1993,wayner1994,gokhale2003,panchamgam2005,panchamgam2006,panchamgam2008,narayanan2011,plawsky2014,akkus2016,akkus2017}

While the kinetic theory of gases \cite{hertz1882,knudsen1950} is widely used to predict the theoretical maximum rate of evaporation, experiments have always calculated smaller heat fluxes than the kinetic limit.\cite{eames1997} However, two recent experimental studies have attracted the attention of scientific community by reporting evaporation fluxes one to two orders of magnitude higher than the prediction of kinetic theory.\cite{radha2016,li2017} These unexpectedly high flux values were attributed to the possible underestimation of evaporation area in the first study,\cite{radha2016} where the stretching of water meniscus over the flat surface adjacent to the channel mouth was speculated. On the other hand, the second study,\cite{li2017} reported evaporation rates from a water meniscus located at a channel entrance defined by four sharp edges, which eliminates the possibility of outstretching of the meniscus. Moreover, extension of meniscus within the channel was also estimated by \cite{li2017} using a model in the literature,\cite{narayanan2011} and the evaporation area was modified accordingly. Regardless of the increased evaporation area, estimated heat fluxes still exceeded the kinetic theory limit, indicating the need for fully understanding the characteristics of an evaporating contact line. Undoubtedly, an accurate calculation of the evaporation flux depends on the precise estimation of effective surface area of evaporation, which is extremely challenging to quantify due to lack of information about the molecular/atomic nature of evaporating contact line.\cite{barkay2013} Thermal scientists have long believed that the adsorbed tail of the evaporating contact line does not possess a net evaporation rate.\cite{wayner1976,dasgupta1993,narayanan2011,plawsky2014} While theoretical studies assumed a balance between evaporation and condensation rates and calculated the equilibrium adsorbed film thickness based on the minimum free energy principle,\cite{narayanan2011} a considerable amount of experimental effort has been devoted to observe the vicinity of the contact line.\cite{dasgupta1993,chen2014convex,deng2015} Most recently, Mehrizi and Wang\cite{mehrizi2018} detected extremely stretched nano-films of thickness 2--6$\unit{nm}$ attached to a water droplet evaporating to its vapor, and 0--10$\unit{nm}$ film attached to a formamide droplet evaporating in air. Morphological distinction between intrinsic meniscus and nano-films was apparent in their all experiments. Therefore, they  defined  the  contact  line  as  the intersection of intrinsic meniscus and the nano-film. \cite{mehrizi2018} Within the water nano-film, they detected a thinning transition film close to the contact line and a near constant thickness adsorbed film attached to the end of transition film. They characterized the transition film as evaporating due to its thickness variation and adsorbed region as non-evaporating due to its near flat profile. Due to the observation of long nano-films beyond the contact line, we consider this experimental work\cite{mehrizi2018} as a clue suggesting a possible area enlargement for the evaporation. However, identification of effective evaporation area, especially in a few nanometers or sub-nanometer interfaces, where adsorbed film, transition film and intrinsic region are intertwined, is impractical for optical and electron microscopy, because of their nano-scale resolution and operating conditions. Therefore, a comprehensive description of evaporation mechanism at this scale is missing.        

\begin{figure*}
\setlength{\belowcaptionskip}{-10pt} 
\includegraphics[width=4.8in]{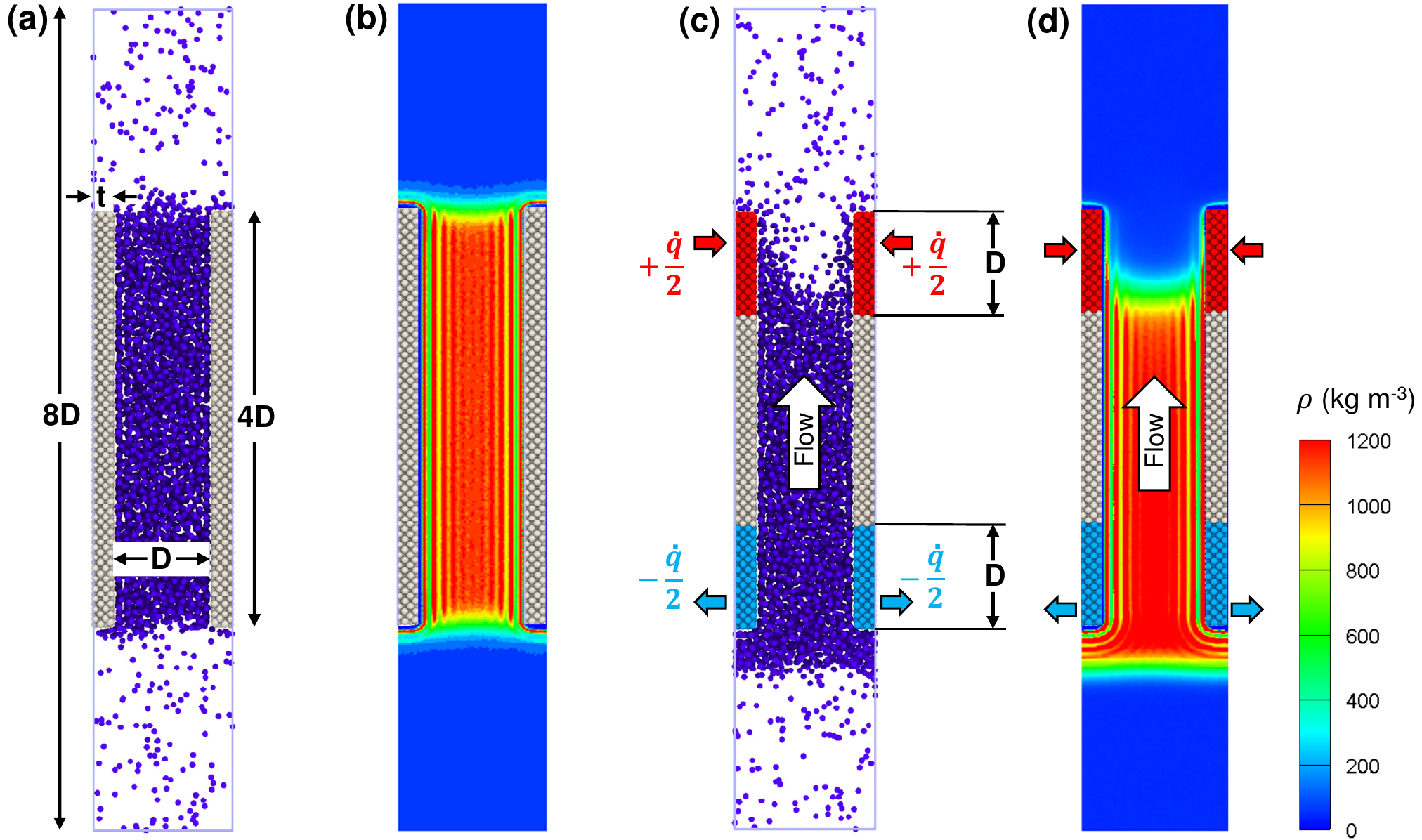}
\caption{\label{fig:fig2} Computational setup. (a) A snapshot of the isothermal configuration. Liquid phase of the saturated Argon mixture (blue spheres) confined between solid Platinum walls (gray spheres). Distance between the channel walls, D, is 3.92 nm and the length of the channel wall and simulation domain is proportional as shown in the figure. Depth of the simulation domain is 3.72 nm. The walls consist of 4 atom layers and have a thickness of t=0.59 nm. See SM for further details about the computational setup and simulation procedure. (b) Density distribution of the isothermal configuration. Two symmetric and statistically stable liquid/vapor interfaces form at the channel inlets. Density layering\cite{heslot1989,cheng2001} is prominent near the walls due to the wall force field effect. (c) A snapshot during the equal energy injection/extraction process. Red and blue parts of the wall show the heating and cooling zones, respectively. The length of these parts is equal to the channel height. The interface at the heating zone is carved due to the evaporative mass loss, while the interface at the cooling zone is expanded due to the condensing fluid atoms. A passive steady liquid flow from the condensing to evaporating interface is initiated due to capillary pumping and molecular diffusion processes.\cite{akkus2018} (d) Density distribution during the equal energy injection/extraction. Both interfaces are statistically stable. The evaporating interface is detached from the channel inlet and receded into the channel. The condensing interface is pushed away from the channel inlet and is nearly flat. Density profiles in (b) and (d) were obtained after time averaging of MD results.}
\end{figure*}

To grasp the physical mechanism of evaporation in capillary nano-conduits, we used molecular dynamics (MD), which is a common tool to study the physical movement of atoms and molecules. Although MD simulations are suitable for nano-scale physical and temporal dimensions, MD results can be successfully used to predict experimentally determined macro-scale system properties.\cite{rahman1964} Evaporation from an interface can be triggered by two basic approaches. First is to make the system sub-saturated by removing vapor and second is to inject energy to the interface molecules by external heating to increase the interface pressure. Modeling of the first mechanism is not practical for the current investigation (see supplementary material (SM), S2). The second mechanism, on the other hand, is realizable as long as the thermodynamic equilibrium of entire system is maintained. To sustain a stable evaporating interface with associated steady-state passive liquid flow, we follow the phase-change driven pumping methodology developed recently.\cite{akkus2018} First, two symmetric isothermal meniscus structures are created within two parallel Platinum walls by condensation of saturated Argon mixture to the liquid phase due to the interaction between fluid and solid wall atoms, whereas vapor phase of Argon occupies rest of the simulation domain \mbox{(Fig.~\ref{fig:fig2}a-b)}. Then, equal energy injection/extraction process is applied to solid atoms in the heating/cooling zones located at the opposite ends of the nanochannel \mbox{(Fig.~\ref{fig:fig2}c-d)}. This approach preserves the thermodynamic state of the mixture by ensuring zero net heat transfer to the system. At the end of 40$\unit{ns}$ simultaneous heating/cooling, an extended meniscus and a flat liquid film (both statistically stationary) evolve at the heated and cooled zones of nanochannel, respectively \mbox{(Fig.~\ref{fig:fig2}d)}. The passive liquid flow through channel is also steady. The methodology and the effect of cutoff density for the interface detection are described in SM.

\begin{figure*}[t!] 
\setlength{\belowcaptionskip}{-10pt} 
\includegraphics[width=6.4in]{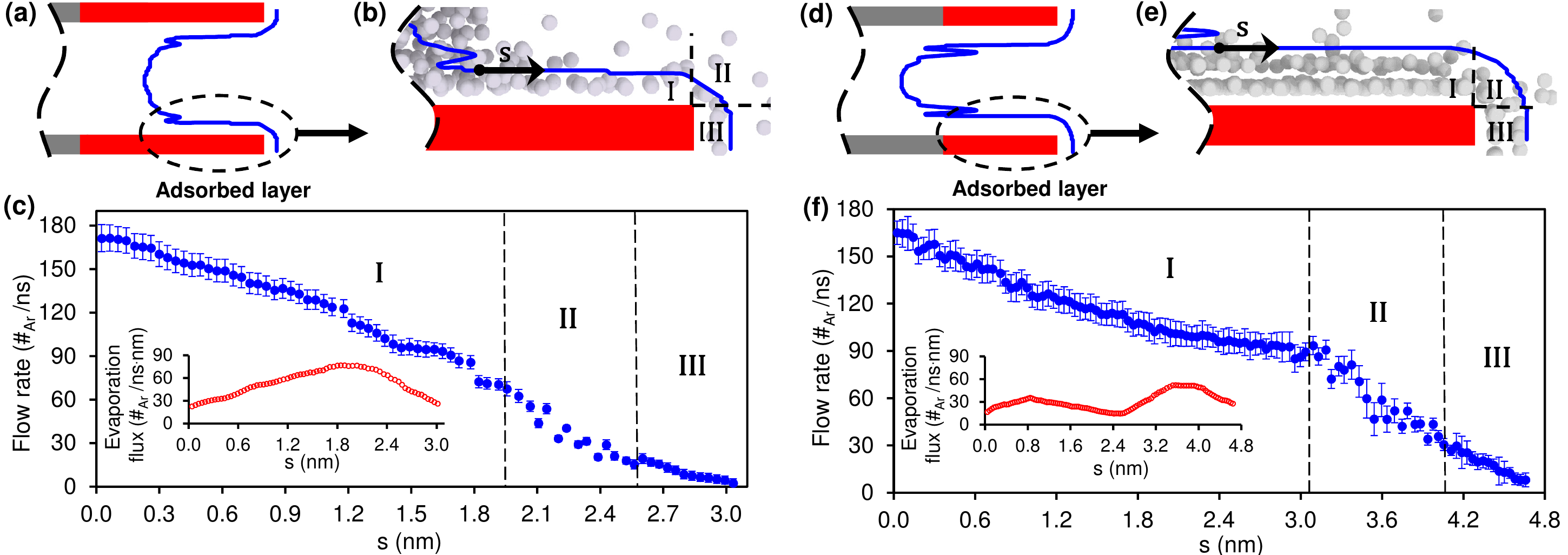}
\caption{\label{fig:fig3} Liquid/vapor interface profiles for (a) the first and (d) second simulations. Constant energy is being injected in the red colored walls. Blue lines show the interface. Menisci receded into the channel and located within (a) and beyond (d) the heating zone. Symmetric adsorbed layers are attached to the menisci at the channel walls. Close-up view of the adsorbed layers for the first (b) and second (e) simulations. A snapshot of atoms are superpositioned on the time-averaged profile of the adsorbed layers, which consist of (b) monolayer of fluid atoms for the first and (e) bilayer of fluid atoms for the second simulation due to the increased fluid-wall binding energy. Surface coordinate, `s', follows the interface of adsorbed layer. Regions I, II and III show the adsorbed layer segments located on the inner channel wall, corner and side wall, respectively. Mass flow rates along the adsorbed layer for (c) the first and (f) second simulations. Mass flow rate decreases along the adsorbed layer and vanishes at the stagnation plane located at the end of region III (see~Fig.~S3 of SM). In the insets, data points show the variation of evaporation flux (per unit depth) along the adsorbed layers. Data is presented after applying moving average with \mbox{20-data points} window size.}
\end{figure*}

Location and morphology of the evaporating meniscus are functions of the heating rate. While a slight heating yields a negligible meniscus deformation, excessive heating results in burnout of the heated wall. During simulations, we applied different heating rates to observe the response of evaporating meniscus. Location and profile of evaporating meniscus corresponding to different heating loads are determined and qualitatively compared with the results of a recent modeling study\cite{lu2015} in \mbox{Fig.~S2} of SM. The profile shown in \mbox{Fig.~\ref{fig:fig3}a} evolves under the highest heating rate just before the burnout of wall.  At this heating rate, liquid meniscus is detached from the channel tips and receded into channel. However, a thin monolayer still covers the surface of the channel at both inside and side walls. We consider this monolayer as the adsorbed layer due to its near flat thickness along the inner wall \mbox{($4.116\pm0.392\unit{\AA}$)} and side wall \mbox{($2.744\pm0.196\unit{\AA}$)} surfaces. Close-up view of adsorbed layer is shown in \mbox{Fig.~\ref{fig:fig3}b}. To investigate the possible momentum transport within the absorbed layer, mass flow is calculated along the surface coordinate, `s' (see SM, S4). Strikingly, an atomic level mass flow is apparent within the absorbed layer as shown in \mbox{Fig.~\ref{fig:fig3}c} due to the solid-liquid surface tension gradient originating from the temperature gradient together with evaporation along the evaporator surface.\cite{sumith2016} The mass flow decreases continuously along the layer and vanishes at the end of the side wall, where actually two opposite molecular streams merge due to the periodic boundary condition (see SM, S8). The inset shows the variation of evaporation flux (per unit depth), which becomes maximum at the corner (region II).

In our computational experiments, we used Argon as the fluid, due to its suitability to be modelled by Lennard-Jones potential with well-defined atomic interactions\cite{barker1968} and its high volatility,  which enables statistical averaging for vapor pressure in contrast to water (see SM, S2). As solid substrate material, we used Platinum due to its empirically defined interaction parameters\cite{foiles1986} and its argon-philic characteristics. It should be noted that the pumping mechanism used in the simulations requires proper wetting of the wall to keep the liquid phase within channel, otherwise condensed liquid slides over walls and leaves the channel, which prevents the investigation of argon-phobic systems. However, the effect of elevated argon-philicity can be easily examined by increasing the interaction strength between fluid and solid atoms. \mbox{Fig.~\ref{fig:fig3}d} shows the profile of evaporating meniscus in the second simulation, where fluid-wall binding energy is 10 times increased. Due to the increased interaction, thickness of absorbed layer is almost doubled (inner wall: \mbox{$7.840\pm0.392\unit{\AA}$}, side wall: \mbox{$6.272\pm0.196\unit{\AA}$}) enabling the formation of a second atomic layer on the first one \mbox{(Fig.~\ref{fig:fig3}e)}. Surprisingly, the highest attainable heating rate before burnout is nearly doubled and the amount of evaporation from the adsorbed layer remains almost constant at $\sim$170 atoms per second \mbox{(Fig.~\ref{fig:fig3}f)}, despite the increased attraction between fluid and wall atoms, which is supposed to prevent evaporation. This interesting observation, indeed, is related to the mobility of adsorbed layer even at the elevated attractive forces (see the video file in SM, S10). Although the first atomic layer (closest to the wall) has low mobility under the effect of wall-force-field, the second layer is subjected to lesser attraction of solid atoms due to increased distance from the wall, and has high mobility similar to the monolayer formed on the wall in previous simulation. Average evaporation flux (see the inset in \mbox{Fig.~\ref{fig:fig3}f}) is lower than the one in the first simulation, since the adsorbed layer is longer while the total evaporating atomic flux is almost same ($\sim$170$\unit{s^{-1}}$).

Our simulations have demonstrated that considerable amount of evaporation can occur at the adsorbed layer (47\% and 64\% of total evaporating mass for our first and second simulations, respectively) and effective evaporation area can be much larger than the liquid meniscus area due to the evaporation from adsorbed layers. To quantify the effect of evaporation area selection on the evaporation flux estimations, we calculate the evaporating mass flux for the first simulation using three different evaporation areas (see SM, S9), which yield: (a) 3311$\unit{kg}\unit{s^{-1}}\unit{m^{-2}}$ for the cross sectional area the between channel walls; (b) 3792$\unit{kg}\unit{s^{-1}}\unit{m^{-2}}$ for the liquid meniscus area based on a curve fit to liquid/vapor interface; (c) 1386$\unit{kg}\unit{s^{-1}}\unit{m^{-2}}$ for the total liquid/vapor interfacial area including the adsorbed layer interface. Moreover, an approximate upper bound for the evaporation is estimated based on Hertz-Knudsen (H-K) equation assuming the evaporation and condensation coefficients as unity (see SM, S6). While the first two area selections result in higher flux values, inclusion of adsorbed layer render the evaporation flux to a smaller value than the estimated upper bound (3235$\unit{kg}\unit{s^{-1}}\unit{m^{-2}}$). Therefore, inclusion of adsorbed layer in the calculation of effective evaporation area can drop the excessive evaporation fluxes reported in recent experiments\cite{radha2016,li2017} below the kinetic limit calculated by these studies. It should be pointed out that precise calculation of kinetic limit depends on the proper selection of empirical parameters (evaporation and condensation coefficients), and these parameters were reported not to be bounded by unity based on the transition probability concept of quantum mechanics, called statistical rate theory (SRT).\cite{persad2016} However, experimental studies were unable to measure the vapor pressure and temperature near the interface, which is necessary to calculate the coefficients in modified H-K relation based on SRT.\cite{persad2016} Our atomistic level modeling, on the other hand, enables the calculation of these coefficients (see SM, S6) and predicts the kinetic limit of the first simulation (10414$\unit{kg}\unit{s^{-1}}\unit{m^{-2}}$) much higher than evaporation flux values.  

Discovery of lateral molecular transport within adsorbed layers requires the questioning of previous modeling attempts both in molecular and continuum levels. In fact, to date, we are unaware of studies which can construct a molecular model for steady-state evaporating meniscus except two studies. \cite{sumith2016,freund2005} While the computational setup of the former\cite{sumith2016} was not suitable to create and examine a planar adsorbed layer, the latter\cite{freund2005} speculated on the existence of a quasi-crystalline region (adsorbed layer) at the end of the evaporating meniscus due to the vanishing liquid flow. However, extinction of liquid flow does not necessarily imply a quasi-crystalline region, since an opposing cross flow (from the image of simulation domain) merges at the side boundary due to the application of periodic boundary condition, which, in fact, renders the side boundary to a stagnation plane. On the other hand, a massive body of literature exists for the theoretical modeling and experimentation of evaporating meniscus in continuum level.\cite{wayner1976,truong1987,wayner1991,dasgupta1993,wayner1994,gokhale2003,panchamgam2005,panchamgam2006,panchamgam2008,narayanan2011,plawsky2014,akkus2016,akkus2017}
In these studies, the concept of disjoining pressure---suppression of the local film pressure due to the strong interaction between wall and liquid atoms---together with the capillary pressure were responsible for the pressure jump at the liquid-vapor interface. The variations of these pressures drove the liquid flow towards the contact line, where a planar adsorbed layer without any liquid flow is attached due to vanishing capillary and disjoining pressure gradients. The absence of lateral momentum transport confirmed the equilibrated adsorbed layer with zero net evaporation in previous studies. However, origin of the strong passive flow adjacent to a solid wall with a substantial temperature gradient is the variation of solid-liquid surface tension, \cite{sumith2016} and the lateral mass flow in a constant thickness layer requires same amount of evaporation to conserve the mass as shown in \mbox{Fig.~\ref{fig:fig3}}. Previous experiments,
\cite{truong1987,wayner1991,dasgupta1993,wayner1994,gokhale2003,panchamgam2005,panchamgam2006,panchamgam2008}
carried out in near isothermal or slight heating conditions, were unable to show the evaporation from adsorbed layers due to the insufficient thermal gradient of the solid substrate.  


In summary, our computational experiments demonstrated that adsorbed liquid films attached to liquid/vapor interfaces are able to turn sharp corners, transport momentum and evaporate. Although the amount of liquid evaporating from adsorbed layers is negligible for macroscopic systems, this amount is comparable to the total evaporation for systems having a few nanometers or sub-nanometer interfaces. Therefore, precise calculation of evaporation rates in these scales requires atomic modeling of liquid/vapor interfaces. Our computational setup allows steady-state evaporating and condensing liquid/vapor interfaces located in capillary nano-conduits, making it a useful tool for investigating phase-change processes at the nano-scale.

\setlength{\parskip}{-10pt}

\begin{acknowledgments}

Y.A. acknowledges the financial support of ASELSAN Inc. under scholarship program for postgraduate studies. Computations were carried out using high performance computing facilities of Center for Scientific Computation at Southern Methodist University.
\end{acknowledgments}

\bibliography{references}

\begin{thebibliography}{38}%
\makeatletter
\providecommand \@ifxundefined [1]{%
 \@ifx{#1\undefined}
}%
\providecommand \@ifnum [1]{%
 \ifnum #1\expandafter \@firstoftwo
 \else \expandafter \@secondoftwo
 \fi
}%
\providecommand \@ifx [1]{%
 \ifx #1\expandafter \@firstoftwo
 \else \expandafter \@secondoftwo
 \fi
}%
\providecommand \natexlab [1]{#1}%
\providecommand \enquote  [1]{``#1''}%
\providecommand \bibnamefont  [1]{#1}%
\providecommand \bibfnamefont [1]{#1}%
\providecommand \citenamefont [1]{#1}%
\providecommand \href@noop [0]{\@secondoftwo}%
\providecommand \href [0]{\begingroup \@sanitize@url \@href}%
\providecommand \@href[1]{\@@startlink{#1}\@@href}%
\providecommand \@@href[1]{\endgroup#1\@@endlink}%
\providecommand \@sanitize@url [0]{\catcode `\\12\catcode `\$12\catcode
  `\&12\catcode `\#12\catcode `\^12\catcode `\_12\catcode `\%12\relax}%
\providecommand \@@startlink[1]{}%
\providecommand \@@endlink[0]{}%
\providecommand \url  [0]{\begingroup\@sanitize@url \@url }%
\providecommand \@url [1]{\endgroup\@href {#1}{\urlprefix }}%
\providecommand \urlprefix  [0]{URL }%
\providecommand \Eprint [0]{\href }%
\providecommand \doibase [0]{http://dx.doi.org/}%
\providecommand \selectlanguage [0]{\@gobble}%
\providecommand \bibinfo  [0]{\@secondoftwo}%
\providecommand \bibfield  [0]{\@secondoftwo}%
\providecommand \translation [1]{[#1]}%
\providecommand \BibitemOpen [0]{}%
\providecommand \bibitemStop [0]{}%
\providecommand \bibitemNoStop [0]{.\EOS\space}%
\providecommand \EOS [0]{\spacefactor3000\relax}%
\providecommand \BibitemShut  [1]{\csname bibitem#1\endcsname}%
\let\auto@bib@innerbib\@empty
\bibitem [{\citenamefont {Wheeler}\ and\ \citenamefont
  {Stroock}(2008)}]{wheeler2008}%
  \BibitemOpen
  \bibfield  {author} {\bibinfo {author} {\bibfnamefont {T.~D.}\ \bibnamefont
  {Wheeler}}\ and\ \bibinfo {author} {\bibfnamefont {A.~D.}\ \bibnamefont
  {Stroock}},\ }\href@noop {} {\bibfield  {journal} {\bibinfo  {journal}
  {Nature}\ }\textbf {\bibinfo {volume} {455}},\ \bibinfo {pages} {208}
  (\bibinfo {year} {2008})}\BibitemShut {NoStop}%
\bibitem [{\citenamefont {Ghasemi}\ \emph {et~al.}(2014)\citenamefont
  {Ghasemi}, \citenamefont {Ni}, \citenamefont {Marconnet}, \citenamefont
  {Loomis}, \citenamefont {Yerci}, \citenamefont {Miljkovic},\ and\
  \citenamefont {Chen}}]{ghasemi2014}%
  \BibitemOpen
  \bibfield  {author} {\bibinfo {author} {\bibfnamefont {H.}~\bibnamefont
  {Ghasemi}}, \bibinfo {author} {\bibfnamefont {G.}~\bibnamefont {Ni}},
  \bibinfo {author} {\bibfnamefont {A.~M.}\ \bibnamefont {Marconnet}}, \bibinfo
  {author} {\bibfnamefont {J.}~\bibnamefont {Loomis}}, \bibinfo {author}
  {\bibfnamefont {S.}~\bibnamefont {Yerci}}, \bibinfo {author} {\bibfnamefont
  {N.}~\bibnamefont {Miljkovic}}, \ and\ \bibinfo {author} {\bibfnamefont
  {G.}~\bibnamefont {Chen}},\ }\href@noop {} {\bibfield  {journal} {\bibinfo
  {journal} {Nat. Commun.}\ }\textbf {\bibinfo {volume} {5}},\ \bibinfo {pages}
  {4449} (\bibinfo {year} {2014})}\BibitemShut {NoStop}%
\bibitem [{\citenamefont {Ni}\ \emph {et~al.}(2016)\citenamefont {Ni},
  \citenamefont {Li}, \citenamefont {Boriskina}, \citenamefont {Li},
  \citenamefont {Yang}, \citenamefont {Zhang},\ and\ \citenamefont
  {Chen}}]{ni2016}%
  \BibitemOpen
  \bibfield  {author} {\bibinfo {author} {\bibfnamefont {G.}~\bibnamefont
  {Ni}}, \bibinfo {author} {\bibfnamefont {G.}~\bibnamefont {Li}}, \bibinfo
  {author} {\bibfnamefont {S.~V.}\ \bibnamefont {Boriskina}}, \bibinfo {author}
  {\bibfnamefont {H.}~\bibnamefont {Li}}, \bibinfo {author} {\bibfnamefont
  {W.}~\bibnamefont {Yang}}, \bibinfo {author} {\bibfnamefont {T.~J.}\
  \bibnamefont {Zhang}}, \ and\ \bibinfo {author} {\bibfnamefont
  {G.}~\bibnamefont {Chen}},\ }\href@noop {} {\bibfield  {journal} {\bibinfo
  {journal} {Nat. Energy}\ }\textbf {\bibinfo {volume} {1}},\ \bibinfo {pages}
  {16126} (\bibinfo {year} {2016})}\BibitemShut {NoStop}%
\bibitem [{\citenamefont {Lee}, \citenamefont {Laoui},\ and\ \citenamefont
  {Karnik}(2014)}]{lee2014}%
  \BibitemOpen
  \bibfield  {author} {\bibinfo {author} {\bibfnamefont {J.}~\bibnamefont
  {Lee}}, \bibinfo {author} {\bibfnamefont {T.}~\bibnamefont {Laoui}}, \ and\
  \bibinfo {author} {\bibfnamefont {R.}~\bibnamefont {Karnik}},\ }\href@noop {}
  {\bibfield  {journal} {\bibinfo  {journal} {Nat. Nanotechnol.}\ }\textbf
  {\bibinfo {volume} {9}},\ \bibinfo {pages} {317} (\bibinfo {year}
  {2014})}\BibitemShut {NoStop}%
\bibitem [{\citenamefont {Lynn}\ and\ \citenamefont {Dandy}(2009)}]{lynn2009}%
  \BibitemOpen
  \bibfield  {author} {\bibinfo {author} {\bibfnamefont {N.~S.}\ \bibnamefont
  {Lynn}}\ and\ \bibinfo {author} {\bibfnamefont {D.~S.}\ \bibnamefont
  {Dandy}},\ }\href@noop {} {\bibfield  {journal} {\bibinfo  {journal} {Lab
  Chip}\ }\textbf {\bibinfo {volume} {9}},\ \bibinfo {pages} {3422} (\bibinfo
  {year} {2009})}\BibitemShut {NoStop}%
\bibitem [{\citenamefont {Li}\ \emph {et~al.}(2012)\citenamefont {Li},
  \citenamefont {Wu}, \citenamefont {Wang}, \citenamefont {Wang}, \citenamefont
  {Liu}, \citenamefont {Zhang}, \citenamefont {Chen}, \citenamefont
  {Peterson},\ and\ \citenamefont {Yang}}]{li2012enhancing}%
  \BibitemOpen
  \bibfield  {author} {\bibinfo {author} {\bibfnamefont {D.}~\bibnamefont
  {Li}}, \bibinfo {author} {\bibfnamefont {G.~S.}\ \bibnamefont {Wu}}, \bibinfo
  {author} {\bibfnamefont {W.}~\bibnamefont {Wang}}, \bibinfo {author}
  {\bibfnamefont {Y.~D.}\ \bibnamefont {Wang}}, \bibinfo {author}
  {\bibfnamefont {D.}~\bibnamefont {Liu}}, \bibinfo {author} {\bibfnamefont
  {D.~C.}\ \bibnamefont {Zhang}}, \bibinfo {author} {\bibfnamefont {Y.~F.}\
  \bibnamefont {Chen}}, \bibinfo {author} {\bibfnamefont {G.~P.}\ \bibnamefont
  {Peterson}}, \ and\ \bibinfo {author} {\bibfnamefont {R.}~\bibnamefont
  {Yang}},\ }\href@noop {} {\bibfield  {journal} {\bibinfo  {journal} {Nano
  Lett.}\ }\textbf {\bibinfo {volume} {12}},\ \bibinfo {pages} {3385} (\bibinfo
  {year} {2012})}\BibitemShut {NoStop}%
\bibitem [{\citenamefont {Wayner}, \citenamefont {Kao},\ and\ \citenamefont
  {LaCroix}(1976)}]{wayner1976}%
  \BibitemOpen
  \bibfield  {author} {\bibinfo {author} {\bibfnamefont {P.~C.}\ \bibnamefont
  {Wayner}}, \bibinfo {author} {\bibfnamefont {Y.~K.}\ \bibnamefont {Kao}}, \
  and\ \bibinfo {author} {\bibfnamefont {L.~V.}\ \bibnamefont {LaCroix}},\
  }\href@noop {} {\bibfield  {journal} {\bibinfo  {journal} {Int. J. Heat Mass
  Tran.}\ }\textbf {\bibinfo {volume} {19}},\ \bibinfo {pages} {487} (\bibinfo
  {year} {1976})}\BibitemShut {NoStop}%
\bibitem [{\citenamefont {Truong}\ and\ \citenamefont
  {Wayner}(1987)}]{truong1987}%
  \BibitemOpen
  \bibfield  {author} {\bibinfo {author} {\bibfnamefont {J.}~\bibnamefont
  {Truong}}\ and\ \bibinfo {author} {\bibfnamefont {P.~C.}\ \bibnamefont
  {Wayner}},\ }\href@noop {} {\bibfield  {journal} {\bibinfo  {journal} {J.
  Chem. Phys.}\ }\textbf {\bibinfo {volume} {87}},\ \bibinfo {pages} {4180}
  (\bibinfo {year} {1987})}\BibitemShut {NoStop}%
\bibitem [{\citenamefont {Sujanani}\ and\ \citenamefont
  {Wayner}(1991)}]{wayner1991}%
  \BibitemOpen
  \bibfield  {author} {\bibinfo {author} {\bibfnamefont {M.}~\bibnamefont
  {Sujanani}}\ and\ \bibinfo {author} {\bibfnamefont {P.~C.}\ \bibnamefont
  {Wayner}},\ }\href@noop {} {\bibfield  {journal} {\bibinfo  {journal} {J.
  Colloid Interf. Sci.}\ }\textbf {\bibinfo {volume} {2}},\ \bibinfo {pages}
  {472} (\bibinfo {year} {1991})}\BibitemShut {NoStop}%
\bibitem [{\citenamefont {DasGupta}, \citenamefont {Schonberg},\ and\
  \citenamefont {Wayner}(1993)}]{dasgupta1993}%
  \BibitemOpen
  \bibfield  {author} {\bibinfo {author} {\bibfnamefont {S.}~\bibnamefont
  {DasGupta}}, \bibinfo {author} {\bibfnamefont {J.}~\bibnamefont {Schonberg}},
  \ and\ \bibinfo {author} {\bibfnamefont {P.}~\bibnamefont {Wayner}},\
  }\href@noop {} {\bibfield  {journal} {\bibinfo  {journal} {J. Heat Transf.}\
  }\textbf {\bibinfo {volume} {115}},\ \bibinfo {pages} {201} (\bibinfo {year}
  {1993})}\BibitemShut {NoStop}%
\bibitem [{\citenamefont {Dasgupta}, \citenamefont {Kim},\ and\ \citenamefont
  {Wayner}(1994)}]{wayner1994}%
  \BibitemOpen
  \bibfield  {author} {\bibinfo {author} {\bibfnamefont {S.}~\bibnamefont
  {Dasgupta}}, \bibinfo {author} {\bibfnamefont {I.}~\bibnamefont {Kim}}, \
  and\ \bibinfo {author} {\bibfnamefont {P.}~\bibnamefont {Wayner}},\
  }\href@noop {} {\bibfield  {journal} {\bibinfo  {journal} {J. Heat Transf.}\
  }\textbf {\bibinfo {volume} {116}},\ \bibinfo {pages} {1007} (\bibinfo {year}
  {1994})}\BibitemShut {NoStop}%
\bibitem [{\citenamefont {Gokhale}, \citenamefont {Plawsky},\ and\
  \citenamefont {Wayner}(2003)}]{gokhale2003}%
  \BibitemOpen
  \bibfield  {author} {\bibinfo {author} {\bibfnamefont {S.}~\bibnamefont
  {Gokhale}}, \bibinfo {author} {\bibfnamefont {J.}~\bibnamefont {Plawsky}}, \
  and\ \bibinfo {author} {\bibfnamefont {P.}~\bibnamefont {Wayner}},\
  }\href@noop {} {\bibfield  {journal} {\bibinfo  {journal} {J. Colloid Interf.
  Sci.}\ }\textbf {\bibinfo {volume} {259}},\ \bibinfo {pages} {354} (\bibinfo
  {year} {2003})}\BibitemShut {NoStop}%
\bibitem [{\citenamefont {Panchamgam}\ \emph {et~al.}(2005)\citenamefont
  {Panchamgam}, \citenamefont {Gokhale}, \citenamefont {Plawsky}, \citenamefont
  {DasGupta},\ and\ \citenamefont {Wayner}}]{panchamgam2005}%
  \BibitemOpen
  \bibfield  {author} {\bibinfo {author} {\bibfnamefont {S.}~\bibnamefont
  {Panchamgam}}, \bibinfo {author} {\bibfnamefont {S.}~\bibnamefont {Gokhale}},
  \bibinfo {author} {\bibfnamefont {J.}~\bibnamefont {Plawsky}}, \bibinfo
  {author} {\bibfnamefont {S.}~\bibnamefont {DasGupta}}, \ and\ \bibinfo
  {author} {\bibfnamefont {P.}~\bibnamefont {Wayner}},\ }\href@noop {}
  {\bibfield  {journal} {\bibinfo  {journal} {J. Heat Transf.}\ }\textbf
  {\bibinfo {volume} {127}},\ \bibinfo {pages} {231} (\bibinfo {year}
  {2005})}\BibitemShut {NoStop}%
\bibitem [{\citenamefont {Panchamgam}, \citenamefont {Plawsky},\ and\
  \citenamefont {Wayner}(2006)}]{panchamgam2006}%
  \BibitemOpen
  \bibfield  {author} {\bibinfo {author} {\bibfnamefont {S.}~\bibnamefont
  {Panchamgam}}, \bibinfo {author} {\bibfnamefont {J.}~\bibnamefont {Plawsky}},
  \ and\ \bibinfo {author} {\bibfnamefont {P.}~\bibnamefont {Wayner}},\
  }\href@noop {} {\bibfield  {journal} {\bibinfo  {journal} {Experimental
  thermal and fluid science}\ }\textbf {\bibinfo {volume} {30}},\ \bibinfo
  {pages} {745} (\bibinfo {year} {2006})}\BibitemShut {NoStop}%
\bibitem [{\citenamefont {Panchamgam}\ \emph {et~al.}(2008)\citenamefont
  {Panchamgam}, \citenamefont {Chatterjee}, \citenamefont {Plawsky},\ and\
  \citenamefont {Wayner}}]{panchamgam2008}%
  \BibitemOpen
  \bibfield  {author} {\bibinfo {author} {\bibfnamefont {S.}~\bibnamefont
  {Panchamgam}}, \bibinfo {author} {\bibfnamefont {A.}~\bibnamefont
  {Chatterjee}}, \bibinfo {author} {\bibfnamefont {J.}~\bibnamefont {Plawsky}},
  \ and\ \bibinfo {author} {\bibfnamefont {P.}~\bibnamefont {Wayner}},\
  }\href@noop {} {\bibfield  {journal} {\bibinfo  {journal} {Int. J. Heat Mass
  Tran.}\ }\textbf {\bibinfo {volume} {51}},\ \bibinfo {pages} {5368} (\bibinfo
  {year} {2008})}\BibitemShut {NoStop}%
\bibitem [{\citenamefont {Narayanan}, \citenamefont {Fedorov},\ and\
  \citenamefont {Joshi}(2011)}]{narayanan2011}%
  \BibitemOpen
  \bibfield  {author} {\bibinfo {author} {\bibfnamefont {S.}~\bibnamefont
  {Narayanan}}, \bibinfo {author} {\bibfnamefont {A.~G.}\ \bibnamefont
  {Fedorov}}, \ and\ \bibinfo {author} {\bibfnamefont {Y.~K.}\ \bibnamefont
  {Joshi}},\ }\href@noop {} {\bibfield  {journal} {\bibinfo  {journal}
  {Langmuir}\ }\textbf {\bibinfo {volume} {27}},\ \bibinfo {pages} {10666}
  (\bibinfo {year} {2011})}\BibitemShut {NoStop}%
\bibitem [{\citenamefont {Plawsky}\ \emph {et~al.}(2014)\citenamefont
  {Plawsky}, \citenamefont {Fedorov}, \citenamefont {Garimella}, \citenamefont
  {Ma}, \citenamefont {Maroo}, \citenamefont {Chen},\ and\ \citenamefont
  {Nam}}]{plawsky2014}%
  \BibitemOpen
  \bibfield  {author} {\bibinfo {author} {\bibfnamefont {J.}~\bibnamefont
  {Plawsky}}, \bibinfo {author} {\bibfnamefont {A.~G.}\ \bibnamefont
  {Fedorov}}, \bibinfo {author} {\bibfnamefont {S.}~\bibnamefont {Garimella}},
  \bibinfo {author} {\bibfnamefont {H.~B.}\ \bibnamefont {Ma}}, \bibinfo
  {author} {\bibfnamefont {S.}~\bibnamefont {Maroo}}, \bibinfo {author}
  {\bibfnamefont {L.}~\bibnamefont {Chen}}, \ and\ \bibinfo {author}
  {\bibfnamefont {Y.}~\bibnamefont {Nam}},\ }\href@noop {} {\bibfield
  {journal} {\bibinfo  {journal} {Nanosc. Microsc. Therm.}\ }\textbf {\bibinfo
  {volume} {18}},\ \bibinfo {pages} {251} (\bibinfo {year} {2014})}\BibitemShut
  {NoStop}%
\bibitem [{\citenamefont {Akku{\c{s}}}\ and\ \citenamefont
  {Dursunkaya}(2016)}]{akkus2016}%
  \BibitemOpen
  \bibfield  {author} {\bibinfo {author} {\bibfnamefont {Y.}~\bibnamefont
  {Akku{\c{s}}}}\ and\ \bibinfo {author} {\bibfnamefont {Z.}~\bibnamefont
  {Dursunkaya}},\ }\href@noop {} {\bibfield  {journal} {\bibinfo  {journal}
  {Int. J. Heat Mass Tran.}\ }\textbf {\bibinfo {volume} {101}},\ \bibinfo
  {pages} {742} (\bibinfo {year} {2016})}\BibitemShut {NoStop}%
\bibitem [{\citenamefont {Akku{\c{s}}}\ \emph {et~al.}(2017)\citenamefont
  {Akku{\c{s}}}, \citenamefont {Tarman}, \citenamefont {{\c{C}}etin},\ and\
  \citenamefont {Dursunkaya}}]{akkus2017}%
  \BibitemOpen
  \bibfield  {author} {\bibinfo {author} {\bibfnamefont {Y.}~\bibnamefont
  {Akku{\c{s}}}}, \bibinfo {author} {\bibfnamefont {H.~I.}\ \bibnamefont
  {Tarman}}, \bibinfo {author} {\bibfnamefont {B.}~\bibnamefont {{\c{C}}etin}},
  \ and\ \bibinfo {author} {\bibfnamefont {Z.}~\bibnamefont {Dursunkaya}},\
  }\href@noop {} {\bibfield  {journal} {\bibinfo  {journal} {Int. J. Therm.
  Sci.}\ }\textbf {\bibinfo {volume} {121}},\ \bibinfo {pages} {237} (\bibinfo
  {year} {2017})}\BibitemShut {NoStop}%
\bibitem [{\citenamefont {Hertz}(1882)}]{hertz1882}%
  \BibitemOpen
  \bibfield  {author} {\bibinfo {author} {\bibfnamefont {H.}~\bibnamefont
  {Hertz}},\ }\href@noop {} {\bibfield  {journal} {\bibinfo  {journal} {Ann.
  Phys.}\ }\textbf {\bibinfo {volume} {253}},\ \bibinfo {pages} {177} (\bibinfo
  {year} {1882})}\BibitemShut {NoStop}%
\bibitem [{\citenamefont {Knudsen}(1950)}]{knudsen1950}%
  \BibitemOpen
  \bibfield  {author} {\bibinfo {author} {\bibfnamefont {M.}~\bibnamefont
  {Knudsen}},\ }\href@noop {} {\emph {\bibinfo {title} {The kinetic theory of
  gases: some modern aspects}}}\ (\bibinfo  {publisher} {Methuen},\ \bibinfo
  {year} {1950})\BibitemShut {NoStop}%
\bibitem [{\citenamefont {Eames}, \citenamefont {Marr},\ and\ \citenamefont
  {Sabir}(1997)}]{eames1997}%
  \BibitemOpen
  \bibfield  {author} {\bibinfo {author} {\bibfnamefont {I.~W.}\ \bibnamefont
  {Eames}}, \bibinfo {author} {\bibfnamefont {N.~J.}\ \bibnamefont {Marr}}, \
  and\ \bibinfo {author} {\bibfnamefont {H.}~\bibnamefont {Sabir}},\
  }\href@noop {} {\bibfield  {journal} {\bibinfo  {journal} {Int. J. Heat Mass
  Tran.}\ }\textbf {\bibinfo {volume} {40}},\ \bibinfo {pages} {2963} (\bibinfo
  {year} {1997})}\BibitemShut {NoStop}%
\bibitem [{\citenamefont {Radha}\ \emph {et~al.}(2016)\citenamefont {Radha},
  \citenamefont {Esfandiar}, \citenamefont {Wang}, \citenamefont {Rooney},
  \citenamefont {Gopinadhan}, \citenamefont {Keerthi}, \citenamefont
  {Mishchenko}, \citenamefont {Janardanan}, \citenamefont {Blake},
  \citenamefont {Fumagalli}, \citenamefont {Lozada-Hidalgo}, \citenamefont
  {Garaj}, \citenamefont {Haigh}, \citenamefont {Grigorieva}, \citenamefont
  {Wu},\ and\ \citenamefont {Geim}}]{radha2016}%
  \BibitemOpen
  \bibfield  {author} {\bibinfo {author} {\bibfnamefont {B.}~\bibnamefont
  {Radha}}, \bibinfo {author} {\bibfnamefont {A.}~\bibnamefont {Esfandiar}},
  \bibinfo {author} {\bibfnamefont {F.~C.}\ \bibnamefont {Wang}}, \bibinfo
  {author} {\bibfnamefont {A.~P.}\ \bibnamefont {Rooney}}, \bibinfo {author}
  {\bibfnamefont {K.}~\bibnamefont {Gopinadhan}}, \bibinfo {author}
  {\bibfnamefont {A.}~\bibnamefont {Keerthi}}, \bibinfo {author} {\bibfnamefont
  {A.}~\bibnamefont {Mishchenko}}, \bibinfo {author} {\bibfnamefont
  {A.}~\bibnamefont {Janardanan}}, \bibinfo {author} {\bibfnamefont
  {P.}~\bibnamefont {Blake}}, \bibinfo {author} {\bibfnamefont
  {L.}~\bibnamefont {Fumagalli}}, \bibinfo {author} {\bibfnamefont
  {M.}~\bibnamefont {Lozada-Hidalgo}}, \bibinfo {author} {\bibfnamefont
  {S.}~\bibnamefont {Garaj}}, \bibinfo {author} {\bibfnamefont {S.~J.}\
  \bibnamefont {Haigh}}, \bibinfo {author} {\bibfnamefont {I.~V.}\ \bibnamefont
  {Grigorieva}}, \bibinfo {author} {\bibfnamefont {H.~A.}\ \bibnamefont {Wu}},
  \ and\ \bibinfo {author} {\bibfnamefont {A.~K.}\ \bibnamefont {Geim}},\
  }\href@noop {} {\bibfield  {journal} {\bibinfo  {journal} {Nature}\ }\textbf
  {\bibinfo {volume} {538}},\ \bibinfo {pages} {222} (\bibinfo {year}
  {2016})}\BibitemShut {NoStop}%
\bibitem [{\citenamefont {Li}\ \emph {et~al.}(2017)\citenamefont {Li},
  \citenamefont {Alibakhshi}, \citenamefont {Zhao},\ and\ \citenamefont
  {Duan}}]{li2017}%
  \BibitemOpen
  \bibfield  {author} {\bibinfo {author} {\bibfnamefont {Y.}~\bibnamefont
  {Li}}, \bibinfo {author} {\bibfnamefont {M.~A.}\ \bibnamefont {Alibakhshi}},
  \bibinfo {author} {\bibfnamefont {Y.}~\bibnamefont {Zhao}}, \ and\ \bibinfo
  {author} {\bibfnamefont {C.}~\bibnamefont {Duan}},\ }\href@noop {} {\bibfield
   {journal} {\bibinfo  {journal} {Nano Lett.}\ }\textbf {\bibinfo {volume}
  {17}},\ \bibinfo {pages} {4813} (\bibinfo {year} {2017})}\BibitemShut
  {NoStop}%
\bibitem [{\citenamefont {Barkay}(2013)}]{barkay2013}%
  \BibitemOpen
  \bibfield  {author} {\bibinfo {author} {\bibfnamefont {Z.}~\bibnamefont
  {Barkay}},\ }in\ \href@noop {} {\emph {\bibinfo {booktitle} {Nanodroplets}}}\
  (\bibinfo  {publisher} {Springer},\ \bibinfo {year} {2013})\ pp.\ \bibinfo
  {pages} {51--72}\BibitemShut {NoStop}%
\bibitem [{\citenamefont {Chen}, \citenamefont {Yu},\ and\ \citenamefont
  {Wang}(2014)}]{chen2014convex}%
  \BibitemOpen
  \bibfield  {author} {\bibinfo {author} {\bibfnamefont {L.}~\bibnamefont
  {Chen}}, \bibinfo {author} {\bibfnamefont {J.}~\bibnamefont {Yu}}, \ and\
  \bibinfo {author} {\bibfnamefont {H.}~\bibnamefont {Wang}},\ }\href@noop {}
  {\bibfield  {journal} {\bibinfo  {journal} {ACS Nano}\ }\textbf {\bibinfo
  {volume} {8}},\ \bibinfo {pages} {11493} (\bibinfo {year}
  {2014})}\BibitemShut {NoStop}%
\bibitem [{\citenamefont {Deng}\ \emph {et~al.}(2015)\citenamefont {Deng},
  \citenamefont {Chen}, \citenamefont {Yu},\ and\ \citenamefont
  {Wang}}]{deng2015}%
  \BibitemOpen
  \bibfield  {author} {\bibinfo {author} {\bibfnamefont {Y.}~\bibnamefont
  {Deng}}, \bibinfo {author} {\bibfnamefont {L.}~\bibnamefont {Chen}}, \bibinfo
  {author} {\bibfnamefont {J.}~\bibnamefont {Yu}}, \ and\ \bibinfo {author}
  {\bibfnamefont {H.}~\bibnamefont {Wang}},\ }\href@noop {} {\bibfield
  {journal} {\bibinfo  {journal} {Int. J. Heat Mass Tran.}\ }\textbf {\bibinfo
  {volume} {91}},\ \bibinfo {pages} {1114} (\bibinfo {year}
  {2015})}\BibitemShut {NoStop}%
\bibitem [{\citenamefont {Mehrizi}\ and\ \citenamefont
  {Wang}(2018)}]{mehrizi2018}%
  \BibitemOpen
  \bibfield  {author} {\bibinfo {author} {\bibfnamefont {A.~A.}\ \bibnamefont
  {Mehrizi}}\ and\ \bibinfo {author} {\bibfnamefont {H.}~\bibnamefont {Wang}},\
  }\href@noop {} {\bibfield  {journal} {\bibinfo  {journal} {Int. J. Heat Mass
  Tran.}\ }\textbf {\bibinfo {volume} {124}},\ \bibinfo {pages} {279} (\bibinfo
  {year} {2018})}\BibitemShut {NoStop}%
\bibitem [{\citenamefont {Rahman}(1964)}]{rahman1964}%
  \BibitemOpen
  \bibfield  {author} {\bibinfo {author} {\bibfnamefont {A.}~\bibnamefont
  {Rahman}},\ }\href@noop {} {\bibfield  {journal} {\bibinfo  {journal} {Phys.
  Rev.}\ }\textbf {\bibinfo {volume} {136}},\ \bibinfo {pages} {A405} (\bibinfo
  {year} {1964})}\BibitemShut {NoStop}%
\bibitem [{\citenamefont {Akkus}\ and\ \citenamefont
  {Beskok}(2018)}]{akkus2018}%
  \BibitemOpen
  \bibfield  {author} {\bibinfo {author} {\bibfnamefont {Y.}~\bibnamefont
  {Akkus}}\ and\ \bibinfo {author} {\bibfnamefont {A.}~\bibnamefont {Beskok}},\
  }\href@noop {} {\bibfield  {journal} {\bibinfo  {journal} {arXiv preprint
  arXiv:1804.06056}\ } (\bibinfo {year} {2018})}\BibitemShut {NoStop}%
\bibitem [{\citenamefont {Heslot}, \citenamefont {Fraysse},\ and\ \citenamefont
  {Cazabat}(1989)}]{heslot1989}%
  \BibitemOpen
  \bibfield  {author} {\bibinfo {author} {\bibfnamefont {F.}~\bibnamefont
  {Heslot}}, \bibinfo {author} {\bibfnamefont {N.}~\bibnamefont {Fraysse}}, \
  and\ \bibinfo {author} {\bibfnamefont {A.~M.}\ \bibnamefont {Cazabat}},\
  }\href@noop {} {\bibfield  {journal} {\bibinfo  {journal} {Nature}\ }\textbf
  {\bibinfo {volume} {338}},\ \bibinfo {pages} {640} (\bibinfo {year}
  {1989})}\BibitemShut {NoStop}%
\bibitem [{\citenamefont {Cheng}\ \emph {et~al.}(2001)\citenamefont {Cheng},
  \citenamefont {Fenter}, \citenamefont {Nagy}, \citenamefont {Schlegel},\ and\
  \citenamefont {Sturchio}}]{cheng2001}%
  \BibitemOpen
  \bibfield  {author} {\bibinfo {author} {\bibfnamefont {L.}~\bibnamefont
  {Cheng}}, \bibinfo {author} {\bibfnamefont {P.}~\bibnamefont {Fenter}},
  \bibinfo {author} {\bibfnamefont {K.~L.}\ \bibnamefont {Nagy}}, \bibinfo
  {author} {\bibfnamefont {M.~L.}\ \bibnamefont {Schlegel}}, \ and\ \bibinfo
  {author} {\bibfnamefont {N.~C.}\ \bibnamefont {Sturchio}},\ }\href@noop {}
  {\bibfield  {journal} {\bibinfo  {journal} {Phys. Rev. Lett.}\ }\textbf
  {\bibinfo {volume} {87}},\ \bibinfo {pages} {156103} (\bibinfo {year}
  {2001})}\BibitemShut {NoStop}%
\bibitem [{\citenamefont {Lu}, \citenamefont {Narayanan},\ and\ \citenamefont
  {Wang}(2015)}]{lu2015}%
  \BibitemOpen
  \bibfield  {author} {\bibinfo {author} {\bibfnamefont {Z.}~\bibnamefont
  {Lu}}, \bibinfo {author} {\bibfnamefont {S.}~\bibnamefont {Narayanan}}, \
  and\ \bibinfo {author} {\bibfnamefont {E.}~\bibnamefont {Wang}},\ }\href@noop
  {} {\bibfield  {journal} {\bibinfo  {journal} {Langmuir}\ }\textbf {\bibinfo
  {volume} {31}},\ \bibinfo {pages} {9817} (\bibinfo {year}
  {2015})}\BibitemShut {NoStop}%
\bibitem [{\citenamefont {Sumith}\ and\ \citenamefont
  {Maroo}(2016)}]{sumith2016}%
  \BibitemOpen
  \bibfield  {author} {\bibinfo {author} {\bibfnamefont {Y.}~\bibnamefont
  {Sumith}}\ and\ \bibinfo {author} {\bibfnamefont {S.}~\bibnamefont {Maroo}},\
  }\href@noop {} {\bibfield  {journal} {\bibinfo  {journal} {Langmuir}\
  }\textbf {\bibinfo {volume} {32}},\ \bibinfo {pages} {8593} (\bibinfo {year}
  {2016})}\BibitemShut {NoStop}%
\bibitem [{\citenamefont {Barker}\ and\ \citenamefont
  {Pompe}(1968)}]{barker1968}%
  \BibitemOpen
  \bibfield  {author} {\bibinfo {author} {\bibfnamefont {J.~A.}\ \bibnamefont
  {Barker}}\ and\ \bibinfo {author} {\bibfnamefont {A.}~\bibnamefont {Pompe}},\
  }\href@noop {} {\bibfield  {journal} {\bibinfo  {journal} {Aust. J. Chem.}\
  }\textbf {\bibinfo {volume} {21}},\ \bibinfo {pages} {1683} (\bibinfo {year}
  {1968})}\BibitemShut {NoStop}%
\bibitem [{\citenamefont {Foiles}, \citenamefont {Baskes},\ and\ \citenamefont
  {Daw}(1986)}]{foiles1986}%
  \BibitemOpen
  \bibfield  {author} {\bibinfo {author} {\bibfnamefont {S.~M.}\ \bibnamefont
  {Foiles}}, \bibinfo {author} {\bibfnamefont {M.~I.}\ \bibnamefont {Baskes}},
  \ and\ \bibinfo {author} {\bibfnamefont {M.~S.}\ \bibnamefont {Daw}},\
  }\href@noop {} {\bibfield  {journal} {\bibinfo  {journal} {Phys. Rev. B}\
  }\textbf {\bibinfo {volume} {33}},\ \bibinfo {pages} {7983} (\bibinfo {year}
  {1986})}\BibitemShut {NoStop}%
\bibitem [{\citenamefont {Persad}\ and\ \citenamefont
  {Ward}(2016)}]{persad2016}%
  \BibitemOpen
  \bibfield  {author} {\bibinfo {author} {\bibfnamefont {A.}~\bibnamefont
  {Persad}}\ and\ \bibinfo {author} {\bibfnamefont {C.}~\bibnamefont {Ward}},\
  }\href@noop {} {\bibfield  {journal} {\bibinfo  {journal} {Chem. Rev.}\
  }\textbf {\bibinfo {volume} {116}},\ \bibinfo {pages} {7727} (\bibinfo {year}
  {2016})}\BibitemShut {NoStop}%
\bibitem [{\citenamefont {Freund}(2005)}]{freund2005}%
  \BibitemOpen
  \bibfield  {author} {\bibinfo {author} {\bibfnamefont {J.~B.}\ \bibnamefont
  {Freund}},\ }\href@noop {} {\bibfield  {journal} {\bibinfo  {journal} {Phys.
  Fluids}\ }\textbf {\bibinfo {volume} {17}},\ \bibinfo {pages} {022104}
  (\bibinfo {year} {2005})}\BibitemShut {NoStop}%
\end{thebibliography}%

\end{document}